\documentstyle[11pt,newpaspdm,twoside,graphicx,psfig]{article}
\def\edcomment#1{\iffalse\marginpar{\raggedright\sl#1\/}\else\relax\fi}
\reversemarginpar

\begin{document}
\title{WIMP direct detection and halo structure}
 \author{Anne M Green}
\affil{Physics Department, Stockholm University, Stockholm, 106 91, SWEDEN}

\begin{abstract}
Weakly Interacting Massive Particle (WIMP) direct detection
experiments are just reaching the sensitivity required to detect
Galactic dark matter in the form of neutralinos (or indeed any stable
weakly interacting particle). Detection strategies and data analyses
are often based on the simplifying assumption of a standard spherical,
isothermal halo model, but observations and numerical simulations
indicate that galaxy halos are in fact triaxial and anisotropic, and
contain substructure. The annual modulation and direction dependence
of the event rate (due to the motion of the Earth) provide the best
prospects of distinguishing WIMP scattering from background events,
however these signals depend sensitively on the local WIMP velocity
distribution. I briefly review the status of WIMP direct detection
experiments before discussing the dependence of the annual modulation
signal on astrophysical input, in particular the structure of the
Milky Way halo, and the possibility that the local WIMP distribution
is not smooth.
\end{abstract}

%\section{WIMP direct detection}
\section{WIMPs a (very) brief introduction}

Any stable weakly interacting massive particle (WIMP) in thermal
equilibrium in the early universe will generically have an interesting
present day density, $\Omega_{{\rm WIMP}} \sim {\cal O} (\Omega_{{\rm
CDM}}) \approx 0.3$. Furthermore supersymmetry provides a natural WIMP
candidate, the lightest supersymmetric particle, the neutralino. There
are basically two methods of detecting WIMPs: indirect detection,
which involves detecting the products of WIMP annihilation ($\gamma$,
$\nu$, $\bar{p}$, $e^+$), and direct detection, which involves
detecting the energy deposited in a detector due to elastic scattering
of WIMPs on the detector nuclei. I will focus on WIMP direct
detection. For a review of particle dark matter see e.g. Bergstr\"om
(2000).

\section{Direct detection signals}

Direct detection experiments are just reaching the sensitivity
required to detect WIMPs. The expected event rates are very small (
${\cal O} (10^{-5} - 10)$ counts ${\rm kg^{-1} day^{-1}}$) and
distinguishing a putative Weakly Interacting Massive Particle (WIMP)
signal from backgrounds, such as neutrons from cosmic-ray induced
muons or natural radioactivity, is crucial. The event rate depends on
the velocity of the detector relative to the Galactic rest frame and
the Earth's motion (as shown in Fig. 1) provides two potential WIMP
smoking guns.  Firstly the event rate is direction dependent, being
greatest in the forward direction (Spergel 1988). Secondly the Earth's
velocity, and hence the event rate, varies annually due to the Earth's
orbit about the Sun (Drukier, Freese \& Spergel 1986). If the local
WIMP velocity distribution is isotropic then the annual modulation is
roughly sinusoidal with a maximum in early June (when the Earth's
speed with respect to the Galactic rest frame is largest) and
amplitude of order a few per-cent.

\begin{figure}
\begin{center}
\includegraphics[angle=0,width=1.0\linewidth]{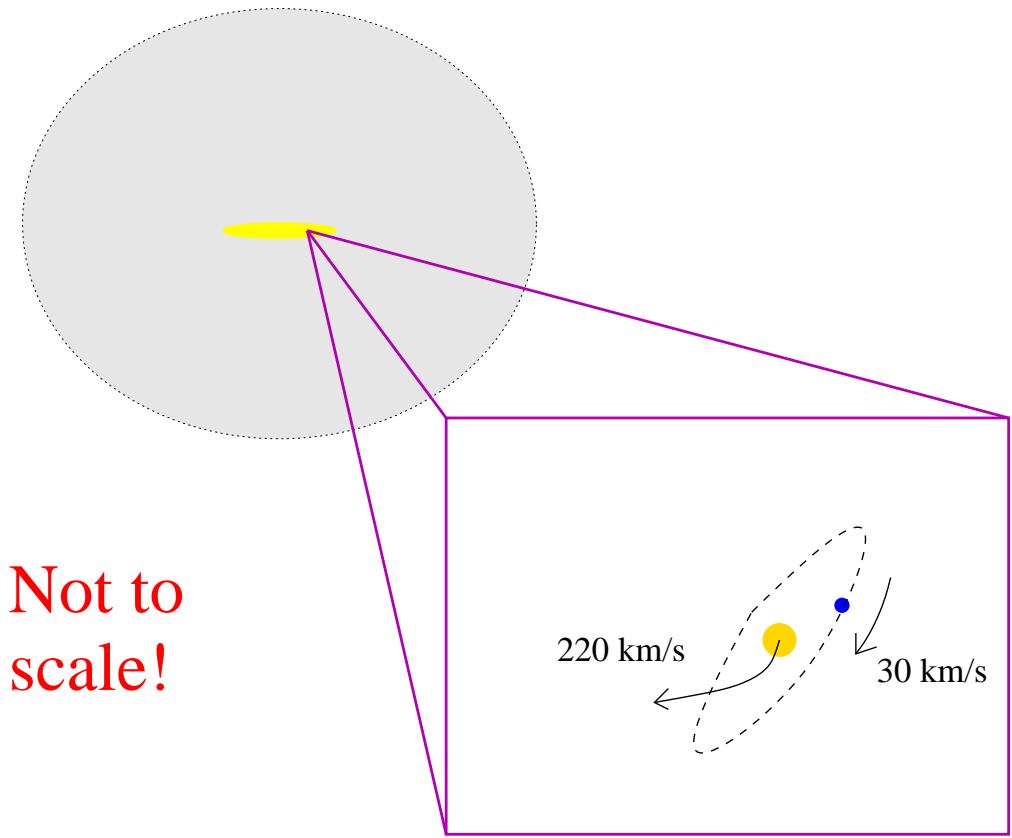}
\end{center}
\caption{A sketch of the motion of the Earth. The Sun moves about the
Galactic centre with circular velocity $v_{{\rm c}} \approx 220 \, \rm
{km \, s^{-1}}$, while the Earth orbits the Sun with speed $v_{{\rm e}}
= 30 \rm{ km \, s^{-1}}$ in a close to circular orbit, inclined at an
angle of roughly $60^{\deg}$ to the Galactic plane.}
\end{figure}

There are currently more than 20 WIMP direct detection experiments
being carried out around the world. I will focus on those currently
producing the most interesting results. The DAMA collaboration, using
NaI with an exposure of 108 000 kg-days at Gran Sasso, have detected
an annual modulation with the properties described above which they
interpret as a WIMP signal (Bernabei et al. 2003). Assuming a standard
halo model with a Maxwellian velocity distribution with dispersion
$\sigma= 270 \rm{km \, s^{-1}}$ (corresponding to an asymptotic halo
circular velocity of $v_{{\rm c}}=220 \, \rm km \, s^{-1}$), they find
a best fit WIMP mass $m_{\chi} \approx 50$ GeV and scattering
cross-section $ \zeta \sigma  \approx 7 \times 10^{-46} \,
{\rm m^2}$ \,\footnote{$\zeta=\rho_{\chi} / (0.3 \, {\rm GeV \, cm^{-3}})
$ parameterizes the uncertainty in the local WIMP density
$\rho_{\chi}$.}. Also of interest are three other experiments using
different targets and a different strategy: Cryogenic Dark Matter
Search (Ge target, 28 kd-days exposure at the Stanford Underground
Facility (Akerib et al. 2003)), Edelweiss (Ge, 32 kg-days
at Modane (Benoit et al. 2002)) and Zeplin I (liquid Xe, 230 kg-days
at Boulby (Barton et al. 2002)). With their smaller exposures these
experiments aim to constrain the mean WIMP scattering rate, rather
than attempting to detect the annual modulation signal.  The full DAMA
allowed region of WIMP mass cross-section parameter space is shown in
fig.2, along with the exclusion limits from the CDMS, Edelweiss and
Zeplin I experiments. The allowed region and the exclusion limits are
all calculated assuming the standard halo model as described above
and, taken at face value, the allowed region appears to be
incompatible with the exclusion limits.

\begin{figure}
\begin{center}
\includegraphics[angle=0,width=0.8\linewidth]{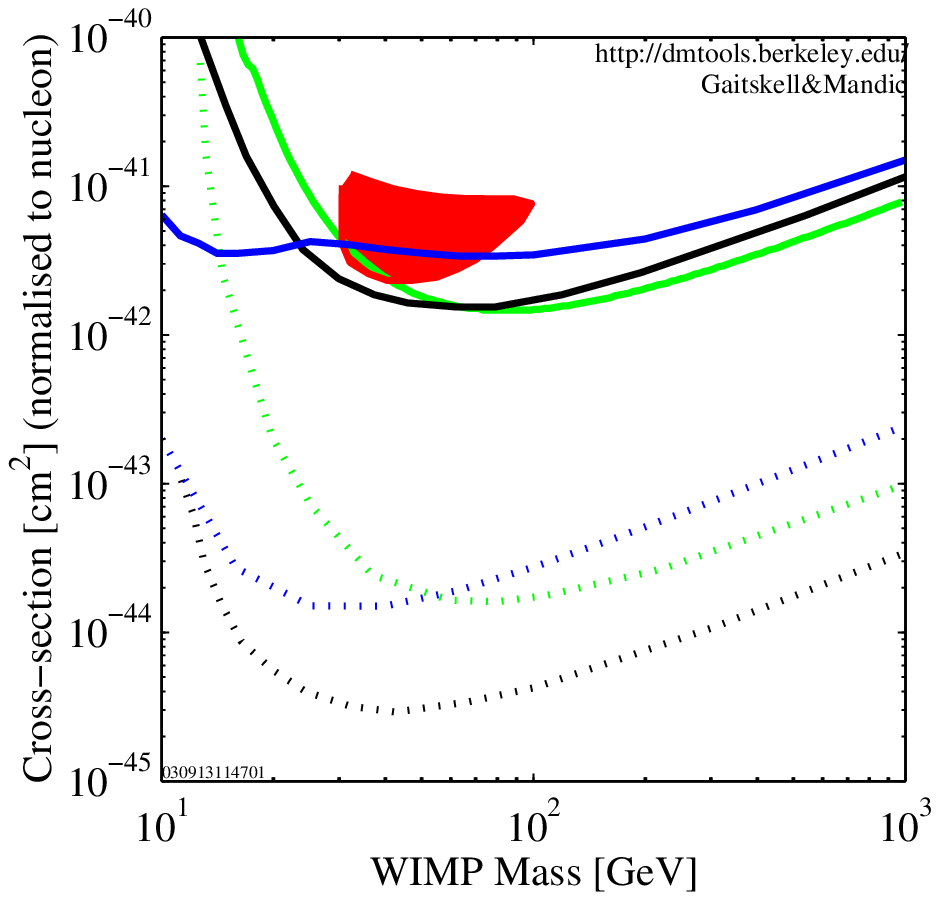}
\includegraphics[angle=0,width=1.0\linewidth]{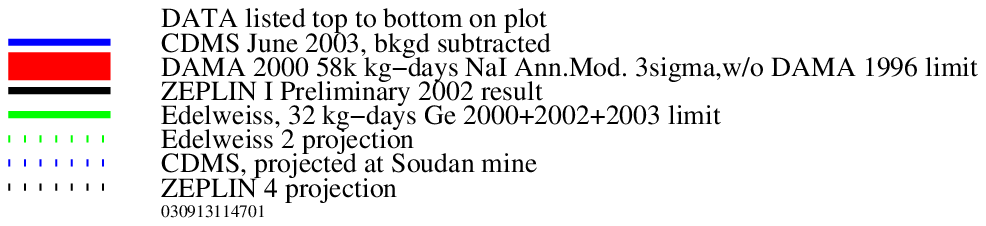}
\end{center}
\caption{Selected experimental results, plotted using the interactive
limit plotter at http://dmtools.berkeley.edu/limitplots, which assumes
the standard halo model. The solid region is the allowed
region corresponding to the DAMA annual modulation signal, the solid
lines the current exclusion limits from the CDMS, Edelweiss and Zeplin
I
experiments and the dotted lines the projected future sensitivities of
these experiments.}
\end{figure}

\section{Local velocity distribution}

The differential elastic scattering rate depends on the local WIMP
density, $\rho_{\chi}$, and the normalised WIMP speed distribution, in
the rest frame of the detector, $f_{v}$, as
\begin{equation}
\frac{{\rm d} R}{{\rm d}E} \propto  \rho_{\chi}
              \int^{\infty}_{v_{{\rm min}}} 
            \frac{f_{v}}{v} \, {\rm d}v \,,  \nonumber
\end{equation}
where $v_{{\rm min}}$ is the minimum WIMP velocity which can
kinematically produce a nuclear recoil of energy $E$.

As discussed above, data analyses usually assume that WIMPs have
an isotropic Maxwellian velocity distribution (i.e. that the Milky Way
halo is an isothermal sphere), but observed and simulated halos are
triaxial, anisotropic and contain substructure.  Exclusion limits,
which depend on the time averaged speed distribution, vary by of
order tens of per-cent when triaxial and anisotropic halo models (with
parameters chosen to match the properties of observed and simulated
halos) are considered, with the shift being experiment dependent
(Green 2002).

The annual modulation signal is far more sensitive to the WIMP
velocity distribution, and Belli et al. (2002) found that considering
``non-standard'' halo models led to a large increase in the size of
the region of $m_{\chi}-\sigma$ parameter space consistent with the
DAMA annual modulation signal. Accurate calculation of the shape and
phase of the annual modulation signal requires all three components of
the Earth's velocity with respect to the Sun, and also the Sun's
motion with respect to the Local Standard of Rest, to be taken into
account (Green 2003). Significantly, if the velocity distribution is
not close to isotropic then the phase and shape of the annual
modulation change and become incompatible with the DAMA annual
modulation signal (Copi \& Krauss 2003, Fornengo \& Scopel 2003, Green
2003).

\section{(very) Small scale structure}

The velocity distributions of analytic halo models are derived via the
Jean's equations, which assume that the phase space distribution
function has reached a steady state. In Cold Dark Matter (CDM)
cosmologies structure forms hierarchically, with galaxy halos forming
from the merger and accretion of smaller subhalos (which themselves
formed from even smaller subhalos), and the local velocity
distribution may not have reached a steady state. Helmi, White \&
Springel (2001) examined the velocity distribution of particles within
a 4 kpc box located 8kpc from the centre of a simulated Milky Way like
halo and found that, apart from a stream of fast moving particles from
a late accreting subhalo, the velocity distribution was well
approximated by a multi-variate gaussian. The WIMP direct detection
event rate, however, depends on the dark matter distribution on
sub-mpc scales, many orders of magnitude smaller than the scales
probed by even the highest resolution galaxy simulations. Moore et
al. (2001) argued that the local velocity distribution will depend
sensitively on the structure and merger history of the first halos to
form, while Stiff \& Widrow (2003) have used a novel reverse technique
to probe the velocity distribution at a single point in a simulation,
finding that it appears to consist of a series of discrete peaks.

The power spectrum on very small scales is an essential input for any
attempt to study the fate of the first subhalos to form and the dark
matter distribution on very small scales. For neutralino CDM
collisional damping and free-streaming erase power on comoving scales
smaller than of order a pc (Hofmann, Schwarz \& St\"ocker 2001, Green,
Hofmann \& Schwarz 2003), and the first subhalos to form have mass of
order $10^{-6} M_{\odot}$ (Green et al. 2003), 12 orders of magnitude
lighter than the smallest subhalos resolved in galaxy
simulations. 

Streams of WIMPs with small velocity dispersion (from late accreting
subhalos which pass through the solar neighbourhood or from the
remnants of the first dense subhalos to form, should they survive to
the present day) will produce steps in the differential event rate,
the position and amplitude of which vary annually.

\section{Summary}

Direct detection of WIMPs would confirm the existence of Cold Dark
Matter (and probe particle physics beyond the standard model).
Accurate astrophysical input (not just the local WIMP velocity
distribution, but also the motion of the detector with respect to the
Galactic rest frame) is required when calculating the WIMP annual
modulation signal. Analyzing data assuming a sinusoidal modulation
with fixed phase could lead to erroneous constraints on, or best fit
values, for the WIMP mass and cross-section, even worse a WIMP signal
could be overlooked. On the other hand using unrealistic halo models
or parameter values could lead to overly restrictive exclusion limits
or a misleadingly large range of allowed values of the WIMP mass and
cross-section. Finally if WIMPs are directly detected then we will be
able to probe the local velocity distribution and perhaps learn about
the (sub-)structure of the Milky Way halo.

\acknowledgments{I would like to thank my collaborators on the
neutralino power spectrum on sub-galactic scales, Stefan Hofmann and
Dominik Schwarz.}


\begin{references}
Akerib, D. S. et al. 2003, hep-ex/0306001 \\
Barton, J. C. et al. 2002, proceedings of the Fourth International Workshop
on the Identification of Dark Matter,  ed. N. J. C. Spooner \&
V. Kudryavtsev (World Scientific), 302 \\
Belli, P., et al. 2002, \prd 66, 043503 \\
Benoit, A. et al. 2002, Phys.Lett.B545, 43 \\
Bergstr\"om, L. 2000, Prog.Phys.63, 793 \\
Bernabei, R. et al. 2003, Riv.N.Cim23, 1 \\
Copi, C. J. \& Krauss, L. M. 2003,  \prd 67, 103507 \\
Drukier, A. K., Freese, K. \& Spergel, D. N. 1986,  \prd 33, 3495 \\ 
Fornengo, N. \& Scopel, S. 2003, astro-ph/0301132 \\
Green, A. M. 2002, \prd 66, 083003 \\
Green, A. M. 2003, \prd 68, 023004 \\
Green, A. M., Hofmann, S. \& Schwarz, D. 2003, astro-ph/0309621 \\
Helmi, A., White, S. D. M. \& Springel, V. 2002, \prd 66, 063502 \\
Hofmann, S., Schwarz, D. J. \& St\"ocker, H. 2001, \prd 64, 083507 \\ 
Moore, B. et al. 2001, \prd 64, 063508 \\
Spergel, D. N. 1988,  \prd 37, 1353 \\
Stiff, D. \& Widrow, L. M. 2003,  \prl 90, 211301 
\end{references}
\end{document}